\documentclass[aps,prl,twocolumn,floatfix,showpacs,superscriptaddress]{revtex4-2}
\usepackage{graphics}
\usepackage{epsfig}
\usepackage{times}
\usepackage{bm}
\usepackage{braket}
\usepackage{color}

\newcommand{\vac}{\left|\mathrm{vac}\right\rangle} 

\usepackage{amsmath,amssymb}	
\begin{document}
\title{Topological Phases in Half-Integer Higher Spin $J_1$-$J_2$ Heisenberg Chains}
\author{Sahinur Reja}
\affiliation{Department of Physics, Jadavpur University, Kolkata 700032,
West Bengal, India}
\author{Satoshi Nishimoto}
\affiliation{Department of Physics, Technical University Dresden, 01069 Dresden, Germany}
\affiliation{Institute for Theoretical Solid State Physics, IFW Dresden, 01069 Dresden, Germany}

\date{\today}

\begin{abstract}
We study the ground state properties of antiferromagnetic $J_1$-$J_2$ chains with half-integer spins ranging from $S=\frac{3}{2}$ to $S=\frac{11}{2}$ using the density-matrix renormalization group method. We map out the ground-state phase diagrams as a function of $\frac{J_2}{J_1}$ containing topological phases with alternating $\frac{2S-1}{2}$ and $\frac{2S+1}{2}$ valence bonds. We identify these topological phases and their boundaries by calculating the string order parameter, the dimer order parameter, and the spin gap for those high-$S$ systems in thermodynamic limit (finite size scaling). We find that these topological regions narrow down inversely with $S$ and converge to a single point at $\frac{J_2}{J_1}=\frac{1}{4}$ in the classical limit -- a critical threshold between commensurate and incommensurate orders. In addition, we extend the discussion of the Majumder-Ghosh state, previously noted only for $S=\frac{1}{2}$, and speculate its possible presence as a ground state in half-integer high spin systems over a substantial range of $\frac{J_2}{J_1}$ values.
\end{abstract}

\maketitle

{\it Introduction. ---}
Topological phases of matter have been a major focus of study since the discovery of the Quantum Hall effect~\cite{PhysRevLett.45.494}, driven by their fundamental physics and potential applications like loss-less power transfer and quantum computing, thanks to their robustness against local perturbations. Interest in this field surged with the discovery of topological insulators in the 2000s, which feature conductive surface states while remaining insulating in the bulk~\cite{RevModPhys.82.3045,RevModPhys.83.1057}. Theoretical advancements, including the development of topological invariants and symmetry-protected topological (SPT) phases~\cite{science.1227224,SPTPhase3_Senthil}, have deepened our understanding of these exotic states. Experimental breakthroughs, such as high-quality sample fabrication and advanced measurement techniques~\cite{science.1173034,zhang2009topological}, have facilitated detailed investigations, paving the way for innovations in spintronics~\cite{moore2010} and quantum computing~\cite{Alicea2012}.

Quantum spin-chain systems with integer spin offer a fertile playground for studying topological phases. A prime example is the $S=1$ Heisenberg chain with nearest-neighbor interactions, which exemplifies an SPT phase~\cite{SPTPhases1_PhysRevB.81.064439,SPTPhase2_PhysRevB.85.075125,SPTPhase3_Senthil}. The ground state of this system is characterized by a gapped valence bond solid (VBS) phase with hidden antiferromagnetic (AFM) order, known as the Haldane phase~\cite{Haldane_1,Haldane_PhysRevLett.50.1153,Affleck_1989}, vividly described by the AKLT model~\cite{AKLT_PhysRevLett.59.799}. Edge states decouple exponentially with chain length in open chains, leading to a four-fold degenerate ground-state manifold~\cite{Tom_kennedy_ED_spin_1_chain}. The phase richness expands with modifications like bond alternation~\cite{bond_alternation_1_PhysRevB.41.9592,bond_alternation_2_PhysRevB.36.5291,bond_alternation_3_doi:10.1143/JPSJ.63.1277,PhysRevB.94.235155}, XXZ anisotropy~\cite{PhysRevB.46.866,PhysRevB.46.13914}, and single-ion anisotropy~\cite{JPSJ.69.237,PhysRevB.67.104401}. The Haldane phase is also realized in integer-spin Heisenberg AFM chains with $S>1$.

The exploration of topological phases, as demonstrated in the $S=1$ Heisenberg chain, reveals significant effects of modifications such as magnetic frustration in higher spin systems. In general, magnetic frustration not only diversifies the quantum phases but also enhances our understanding of their topological properties~\cite{Roth_U.Schollwock_PhysRevB.58.9264,sequential_phase_transition_S234_PhysRevB.100.014438}. A simple realization of a frustrated spin model is the Heisenberg spin chain with nearest and next-nearest neighbor interactions, known as the so-called $J_1-J_2$ spin-chain. This model allows us to emphasize how lattice geometry, interaction details, and spin magnitude affect topological robustness. For example, integer spin ($S=2$, $3$, and $4$) $J_1-J_2$ model has been investigated using Berry phase analysis to show sequential transitions between topological phases~\cite{sequential_phase_transition_S234_PhysRevB.100.014438}. However, these transitions are understood from the perspective of energetically favorable VBS states rather than the effect of magnetic frustration.

On the contrary, for half-integer spin $J_1$-$J_2$ chains, the effects of magnetic frustration are more pronounced, making the problem more challenging. For $S=\frac{3}{2}$, earlier density-matrix renormalization group (DMRG) studies~\cite{PhysRevB.63.174430, Roth_U.Schollwock_PhysRevB.58.9264} laid the groundwork, and recent research~\cite{Frederic_Mila_Spin3_2_PhysRevB.101.174407} identifies four phases as a function of $\frac{J_2}{J_1}$: a Tomonaga-Luttinger-liquid (TLL) phase for $\frac{J_2}{J_1} \lesssim 0.31$, a partially dimerized (PD) gapped phase for $0.31 \lesssim \frac{J_2}{J_1} \lesssim 0.52$, a critical floating phase for $0.52 \lesssim \frac{J_2}{J_1} \lesssim 1.20$, and a fully dimerized (FD) phase for $\frac{J_2}{J_1} \gtrsim 1.20$. Despite the quantitative differences, similar phase diagrams have been observed for $S=\frac{5}{2}$~\cite{Frederic_Mila_Spin5_2_PhysRevB.105.174402}. However, the evolution of the topological phases arising in these quantum models and how these phases connect to the classical limit with increasing spin remains a non-trivial and open question. To answer these questions, we revisit the $S=\frac{3}{2}$ and $S=\frac{5}{2}$ $J_1$-$J_2$ systems and extend our study upto $S=\frac{11}{2}$.

In this Letter, we study the ground states of AFM $J_1$-$J_2$ chain with half-integer spins ranging from $S=\frac{3}{2}$ to $S=\frac{11}{2}$, using the DMRG method. We calculate the dimer order parameter, spin gap, and string order parameter in the thermodynamic limit for a wide range of $\frac{J_2}{J_1}$ to obtain the phase diagrams. We find that the range of PD phase narrows down with $1/S$ approximately, and converges to a single point at $\frac{J_2}{J_1}=\frac{1}{4}$ in the classical limit ($S\rightarrow\infty$). We also extend the discussion of the Majumder-Ghosh (MG) state, previously noted only for $S=\frac{1}{2}$, showing its potential presence as a ground state in higher spin systems over a substantial range of $\frac{J_2}{J_1}$ values.

{\it Model and method. ---}
The Hamiltonian of the AFM $J_1$-$J_2$ chain is given by:
\begin{align}
	{\cal H} = J_1\sum_{i}\mathbf{S}_i\cdot\mathbf{S}_{i+1}+J_2\sum_{i}\mathbf{S}_i\cdot\mathbf{S}_{i+2},
	\label{eq:hamiltonian}
\end{align}
where $\mathbf{S}_i$ is the spin operator at site $i$, interacting with nearest and next-nearest neighbor spins with coupling strengths $J_1\,(>0)$ and $J_2\,(>0)$, respectively. Here we fix $J_1=1$ as the unit of energy hereafter.

\begin{figure}[tbh]
	\centering
	\includegraphics[width=0.8\linewidth]{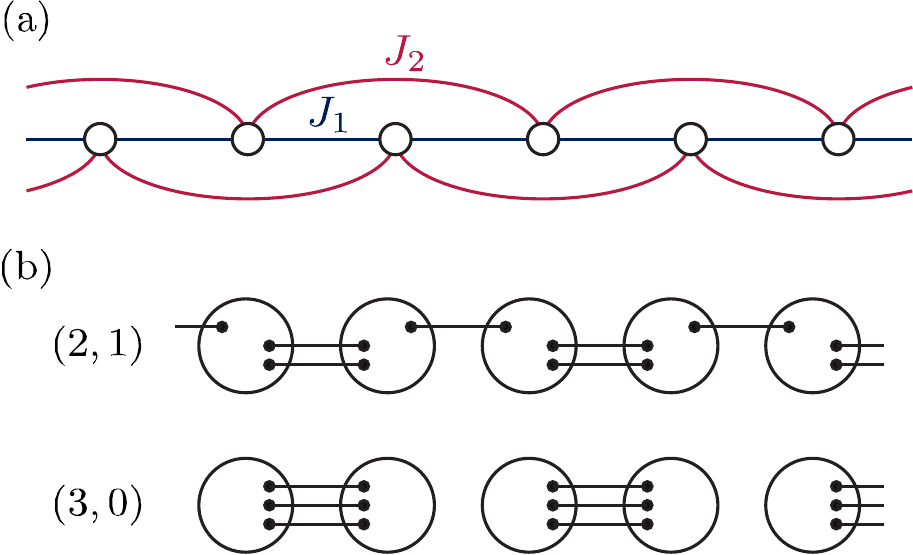}
	\caption{
		(a) Lattice structure of spin-$S$ $J_1$-$J_2$ Heisenberg chain
		where each empty circle denotes a spin-$S$ site.
		(b) Schematic illustrations of dimerized VBS states for an
		$S=\frac{3}{2}$ system, depicting the partially dimerized (2,1)
		phase and the fully dimerized (3,0) phase. Each link represents
		a spin-$\frac{1}{2}$ singlet bond.
	}
	\label{fig:lattice}
\end{figure}

We employ the DMRG method~\cite{White1992} to explore the model described in Eq.~\ref{eq:hamiltonian} for spin magnitudes $S=\frac{3}{2}, \dots, \frac{11}{2}$. Our simulations are conducted under open boundary conditions (OBC), modifying the spin values at the chain edges (unless specified otherwise). For instance, in the $S=\frac{3}{2}$ systems, we test three kinds of edge spins - $S=\frac{1}{2}$, $1$, and $\frac{3}{2}$ - and calculate the relevant physical quantities for each case. We simulate systems up to $L=180$ sites with up to $\chi=5000$ basis states retained in the density matrix, achieving the largest discarded weight $w_{\rm d}=4.4\times10^{-8}$. For spins larger than $S=\frac{3}{2}$, the maximum system sizes, number of retained eigenstates of the density matrix, and discarded weights for these simulations are $(L,\chi,w_{\rm d})=(128,3000,1.6 \times 10^{-7})$ for $S=\frac{5}{2}$, $(L,\chi,w_{\rm d})=(96,3000,4.0 \times 10^{-8})$ for $S=\frac{7}{2}$, $(L,\chi,w_{\rm d})=(64,3800,3.9 \times 10^{-9})$ for $S=\frac{9}{2}$, and $(L,\chi,w_{\rm d})=(40,3400,1.1 \times 10^{-9})$ for $S=\frac{11}{2}$. All numerical results are extrapolated to the thermodynamic limit to ensure broad applicability of our findings.

To describe the dimerized VBS state, we introduce the $(m,n)$-type Schwinger-boson representation as follows~\cite{KTotsuka1995,MNakamura2002}:
\begin{align}
	\ket{(m,n)}=\frac{1}{\sqrt{\mathcal{N}}}
	\prod_{j=1}\bigl(B^{\dagger}_{2j-1,2j}\bigr)^{m}
	\bigl(B^{\dagger}_{2j,2j+1}\bigr)^{n}\vac,
	\label{eq:(m,n)-VBS}
\end{align}
where $B^{\dagger}_{i,j} = a^{\dagger}_{i} b^{\dagger}_{j} - b^{\dagger}_{i} a^{\dagger}_{j}$ involves the bosonic operators $a^{\dagger}_{i}$ and $b^{\dagger}_{i}$, $\mathcal{N}$ is the normalization factor, and $\vac$ denotes the boson vacuum. In this representation, the spin operators are defined by the Schwinger-boson expressions: $S^{+}_{i} = a^{\dagger}_{i} b_{i}$, $S^{-}_{i} = b^{\dagger}_{i} a_{i}$, and $S^{z}_{i} = \frac{1}{2}(a^{\dagger}_{i} a_{i} - b^{\dagger}_{i} b_{i})$. The integers $m$ and $n$ satisfy $m+n=2S$. It is noteworthy that in half-integer spin systems, the $(m,n)$ and $(n,m)$ states are degenerate for a periodic chain, though this degeneracy can be lifted by adjusting the edge spins under OBC.

{\it Majumdar–Ghosh model. ---}
For $S=\frac{1}{2}$, due to its simplicity and the possibility of experimental realizations, the model~\eqref{eq:hamiltonian} has been extensively studied~\cite{j1j2book}, and the ground state is well known across the entire parameter region. At $J_2 < 0.2411$, the system is in a Tomonaga-Luttinger liquid (TLL) phase, and at $J_2 > 0.2411$, it transitions into a spontaneously dimerized VBS phase~\cite{spin12_j1_j2_0.2411_OKAMOTO1992433}. One of the most interesting aspects is that within this dimerized VBS phase, at the specific point $J_2=\frac{1}{2}$, known as the Majumdar–Ghosh (MG) point, the ground state of the system is given analytically and exactly~\cite{majumdar_ghosh_10.1063/1.1664978}. This discussion can be extended to general $S$ as follows. The Hamiltonian~\eqref{eq:hamiltonian} can be reformulated as: ${\cal H} = \frac{1}{4}\sum_{i}[(\mathbf{S}_i+\mathbf{S}_{i+1}+\mathbf{S}_{i+2})^2-(\mathbf{S}_i^2+\mathbf{S}_{i+1}^2+\mathbf{S}_{i+2}^2)]$. For a spin-$S$ system, the total energy achieves its minimum when $(\mathbf{S}_i+\mathbf{S}_{i+1}+\mathbf{S}_{i+2})^2=S(S+1)$, resulting in the ground state energy: $E_{\rm MG}=-\frac{1}{2}S(S+1)L$, where $L$ denotes the number of lattice sites. Although for $S > \frac{1}{2}$, this state is not the ground state, it potentially connects to the ground state for a large region of $J_2$ for half-integer $S$, as demonstrated later.

{\it Physical quantities. ---}
The spontaneous dimerization transition can be directly detected by calculating the dimer order parameter. This parameter is quantitatively assessed by the difference in spin-spin correlations between neighboring bonds, defined as:
\begin{align}
\mathcal{D}_{S_{\rm edge}}= \lim_{L \to \infty} \left| \langle \mathbf{S}_{i-1} \cdot \mathbf{S}_{i} \rangle - \langle \mathbf{S}_{i} \cdot \mathbf{S}_{i+1} \rangle \right|,
\end{align}
where the site $i$ is chosen to be the middle site of the system, i.e., $i=\frac{L}{2}$. Non-zero ${D}_{S_{\rm edge}}$ in thermodynamic limit effectively captures the emergence of long-range dimer order, highlighting the critical point of the transition.

The presence of the dimerized VBS phase can be further confirmed by examining the behavior of correlation functions similar to the string order originally introduced for spin-1 systems. The VBS state exemplifies a state with topological and hidden non-local order. Thus, even in half-integer spin systems, one can define a string order parameter, particularly in a generalized VBS context where the spins form pairs
(dimers)~\cite{nonLocalString_Order_PhysRevB.40.4709,Oshikawa1992HiddenZS}:
\begin{align}
	\mathcal{S}_{S_{\rm edge}}=\lim_{k-j \to \infty}
	\left\langle
	S^{z}_{j}\exp{\Bigl(i\pi\sum^{k-1}_{l=j}S^{z}_{l}\Bigr)}S^{z}_{k}
	\right\rangle.
	\label{eq:Stringorder}
\end{align}
This order parameter captures the essence of the non-local correlations characteristic of VBS states which are not detectable through conventional order parameters. However, its definition depends on the positions of the strong and weak bonds. We calculate this quantity for various edge spins using open chains. Universally, Eq.~\eqref{eq:Stringorder} attains a finite value when there are string bonds between sites $j-1$, $j$ and $k-1$, $k$. Specifically, if an even number of strong bonds lie between these two strong bonds, then $\mathcal{S}_{S_{\rm edge}} < 0$; if an odd number of strong bonds lie between them, $\mathcal{S}_{S_{\rm edge}} > 0$. With larger $S$, the convergence of the string order parameter with respect to $k-j$ becomes slower due to the larger correlation length.

Further, calculating the singlet-triplet spin gap is crucial for understanding the dimerized VBS state because its behavior depends on the kind of edge spin. This gap signals the presence of a gapped, non-magnetic phase, helps characterize the nature of dimerization and singlet structures, and differentiates the VBS phase from critical magnetic phases. It is defined as the energy difference between the lowest-energy triplet state and the singlet ground state:
\begin{align}
	\Delta_{S_{\rm edge}}=\lim_{L \to \infty} [E_0(1)-E_0(0)],
	\label{eq:spingap}
\end{align}
where $E_0(S^z)$ is the lowest energy for a total spin $S^z$ sector. Changes in the gap due to edge conditions can reveal the dimerization pattern and whether the edge spins form effective singlets or remain free, indicating the nature of the VBS and possible critical behavior.

\begin{figure}[t]
	\centering
	\includegraphics[width=0.9\linewidth]{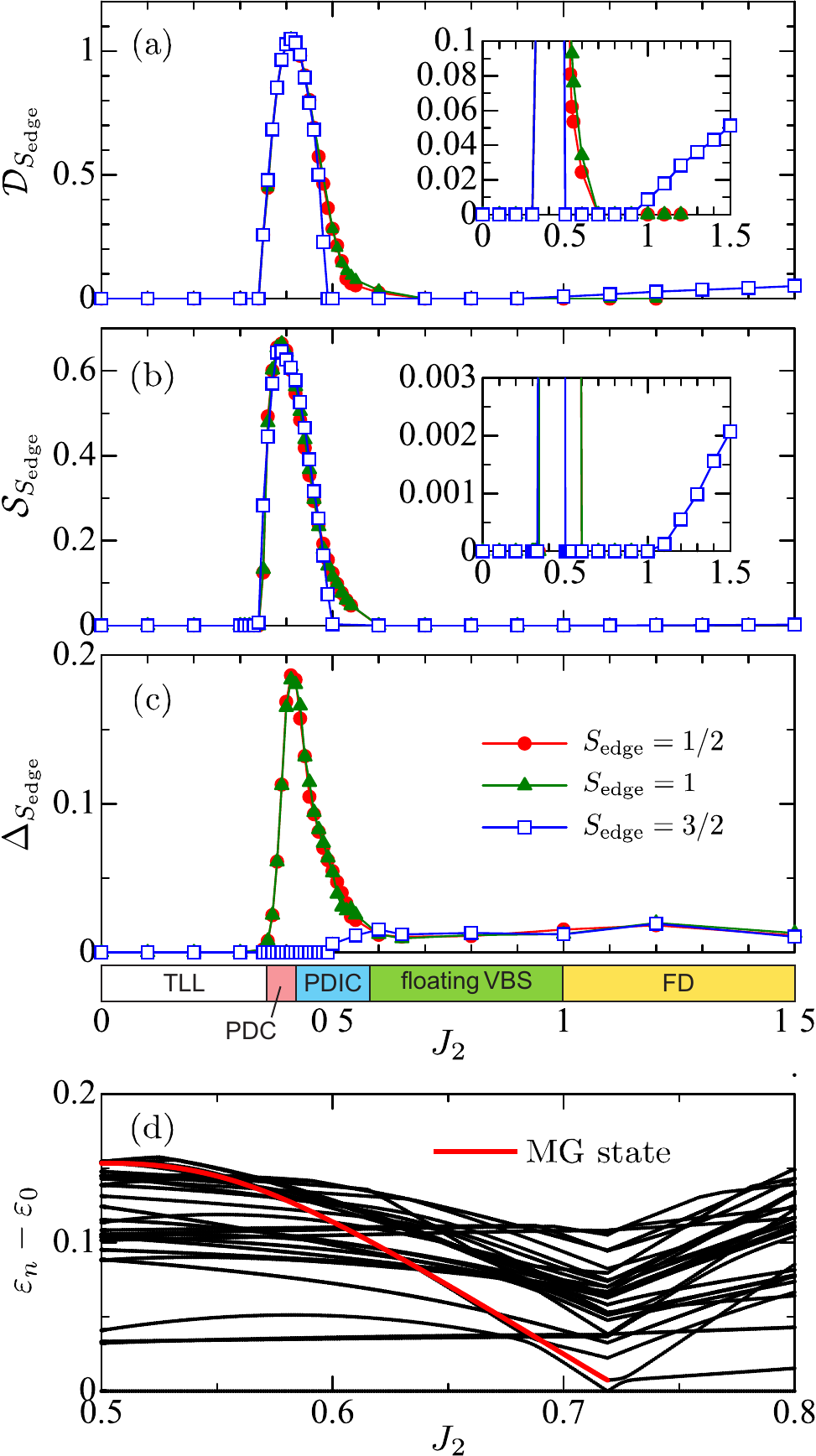}
	\caption{
		Extrapolated values of (a) the dimer order parameter ${\cal D}_{S_{\rm edge}}$, (b) the string order parameter ${\cal S}_{S_{\rm edge}}$, and (c) the spin gap $\Delta_{S_{\rm edge}}$ as a function of $J_2$ for the $S=\frac{3}{2}$ system, considering edge spins $S=\frac{1}{2}$, $1$, and $S=\frac{3}{2}$. The corresponding phases are also indicated. Insets of (b) and (c) show magnified views of the low-value regions. (d) Energy spectrum of the lowest 45 states for an 8-site $S=\frac{3}{2}$ periodic chain, with the MG state highlighted in red.
	}
	\label{fig:S32}
\end{figure}

{\it Results for $S=\frac{3}{2}$. ---}
Let us first examine the results for the case of $S=\frac{3}{2}$. As mentioned earlier, this case has already been investigated in Ref.~\onlinecite{Roth_U.Schollwock_PhysRevB.58.9264,Frederic_Mila_Spin3_2_PhysRevB.101.174407}. Our analysis here serves as a benchmark for these results. In Fig.~\ref{fig:S32}(a-c), we plot the extrapolated values of the dimer order parameter $\mathcal{D}_{S_{\rm edge}}$, the string order parameter $\mathcal{S}_{S_{\rm edge}}$, and the spin gap $\Delta_{S_{\rm edge}}$ as functions of $J_2$ for edge spins $S_{\rm edge}$ of $\frac{1}{2}$, $1$, and $\frac{3}{2}$ as indicated.

Within the range $0.34 \lesssim J_2 \lesssim 0.58$, all three physical quantities are finite for $S_{\rm edge}=\frac{1}{2}$ and $1$. This suggests the formation of a dimerized VBS state with topological hidden order in this region. However, for $S_{\rm edge} = \frac{3}{2}$, the behavior of these quantities differs. Specifically, for $0.34 \lesssim J_2 \lesssim 0.50$, the spin gap is zero, indicating the presence of edge states and suggesting that the VBS state is of the (2,1)- or (1,2)-type, characteristic of a PD phase. We note that this VBS state is of the (1,2)-type for $S_{\rm edge}=\frac{1}{2}$ and of the (2,1)-type for $S_{\rm edge}=1$ with an explicit breaking of translation symmetry in our open chains. In the remaining range $0.50 \lesssim J_2 \lesssim 0.58$, both the dimer and string order parameters vanish, while the spin gap remains finite. These observations are consistent with the disappearance of edge states, as noted in Ref.~\cite{Frederic_Mila_Spin3_2_PhysRevB.101.174407}. For $S_{\rm edge}=\frac{3}{2}$, the explicit breaking of translation symmetry is absent in this region, leading to zero values for the dimer and string order parameters. Concurrently, the vanishing of edge states, coupled with finite-size effects under OBC, allows for the formation of a special VBS structure at the ends of an open chain, thereby giving rise to a finite spin gap. However, it is crucial to note that this spin gap primarily reflects excitations near the edges and differs from the bulk limit. In the bulk limit, the range $0.34 \lesssim J_2 \lesssim 0.58$ would correspond to a PD phase. This illustrates that the absence of edge states in finite systems does not necessarily imply the absence of topological order in infinite systems. The PD phase can be further categorized into commensurate (PDC) and incommensurate (PDIC) types, as differentiated by the presence or absence of oscillations in the size-dependence of physical quantities~\cite{Frederic_Mila_Spin3_2_PhysRevB.101.174407}. Our analysis confirms this classification based on the oscillatory behavior in the size dependence of the measured quantities (see the Supplemental Material).

In the regime $0.58 \lesssim J_2 \lesssim 1.0$, both the dimer and string order parameters vanish, yet a small but finite spin gap exists. This observation seems to be consistent with the floating VBS picture proposed by Ref.~\cite{Frederic_Mila_Spin3_2_PhysRevB.101.174407}. As $J_2$ increases further, near $J_2=1$, the dimer and string order parameters begin to increase. The magnitude of the spin gap appears to be independent of edge spins and reaches its maximum around $J_2=1.2$. This indicates that the region $J_2>1$ corresponds to the FD phase characterized by a (3,0)-type VBS state. Due to the intrinsic incommensurability of the system, estimating the spin gaps for the floating VBS and FD phases is challenging, necessitating careful finite-size scaling analysis. A more detailed analysis is presented in the Supplemental Material.

To further substantiate this, we examined whether the MG eigenstate, which is adiabatically connected to the (3,0)-type VBS state, can be the ground state over a broad range of $J_2$ using a periodic chain with $L=8$. In Fig.~\ref{fig:S32}(d), we plot the energy spectrum relative to the ground state as a function of $J_2$. At $J_2=\frac{1}{2}$, the exact MG eigenstate appears as the 39th excited state. As $J_2$ increases, its energy decreases relative to other eigenstates. Although finite-size effects preclude definitive conclusions, the eigenstate nearly degenerates with the ground state around $J_2\sim 0.72$.

\begin{figure}[t]
\centering
\includegraphics[width=0.8\linewidth]{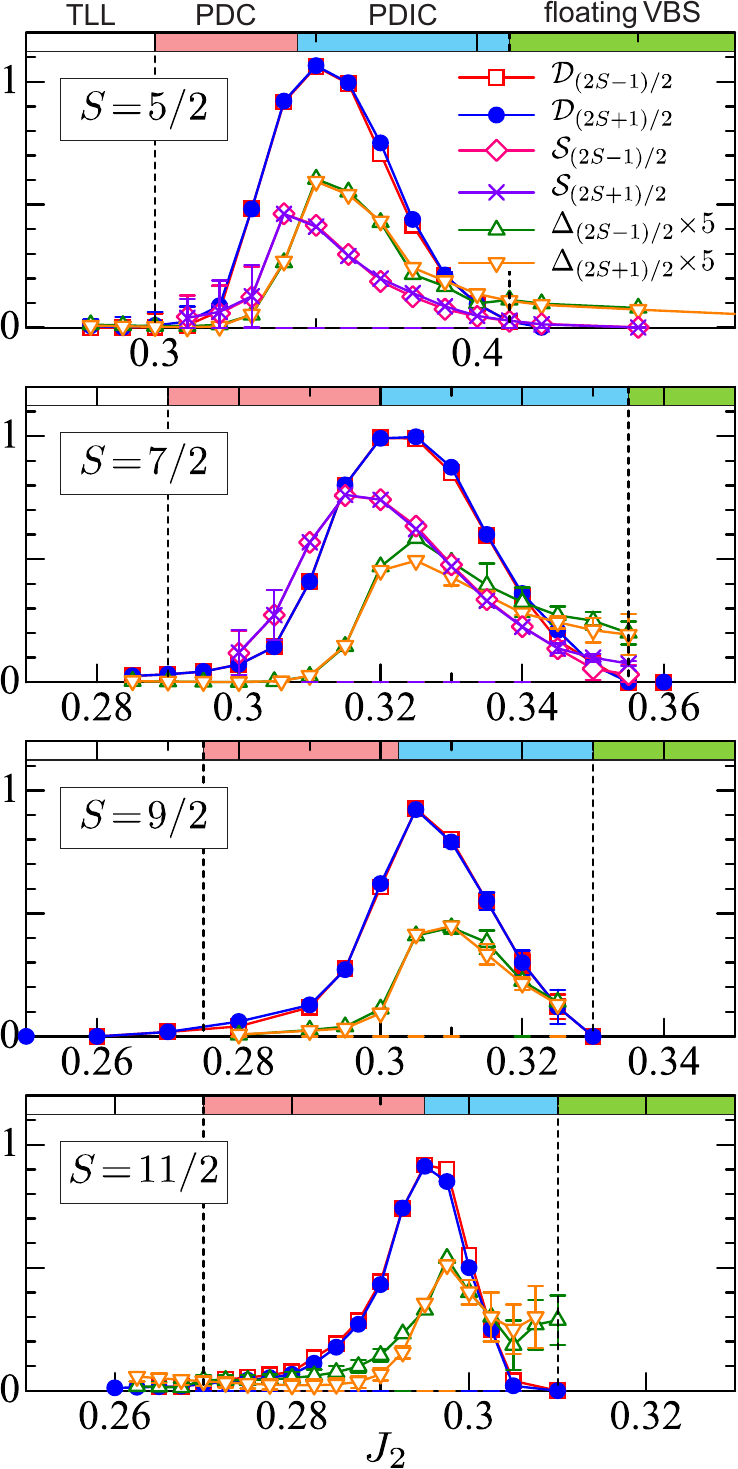}
\caption{
		Extrapolated values of the dimer order parameter ${\cal D}_{S_{\rm edge}}$, the string order parameter ${\cal S}_{S_{\rm edge}}$, the spin gap $\Delta_{S_{\rm edge}}$ as a function of $J_2$ for the $S=\frac{5}{2}$, $S=\frac{7}{2}$, $S=\frac{9}{2}$, and $S=\frac{11}{2}$ systems, considering edge spins $S=\frac{2S-1}{2}$ and $S=\frac{2S+1}{2}$. Results for the string order parameter for $S=\frac{9}{2}$ and $S=\frac{11}{2}$ systems are not shown (see text). The corresponding phases are also indicated.
}
\label{fig:OP}
\end{figure}

{\it Results for higher spins. ---}
Next, we examine the case for $S > \frac{3}{2}$. In Fig.~\ref{fig:OP}, the extrapolated values of the dimer order parameter $\mathcal{D}_{S_{\rm edge}}$, the string order parameter $\mathcal{S}_{S_{\rm edge}}$, and the spin gap $\Delta_{S_{\rm edge}}$ are plotted as functions of $J_2$ for $S = \frac{5}{2}$, $\frac{7}{2}$, $\frac{9}{2}$, and $\frac{11}{2}$ systems. Here, we present results for $S_{\rm edge}=\frac{2S-1}{2}$ and $\frac{2S+1}{2}$. Note that it is hard to estimate $\mathcal{S}_{S_{\rm edge}}$ in the thermodynamic limit for $S=\frac{9}{2}$ and $\frac{11}{2}$ systems due to the large correlation length. The regions where the dimer order parameter is finite are estimated to be:
$0.30 \lesssim J_2 \lesssim 0.42$ for $S=\frac{5}{2}$,
$0.29 \lesssim J_2 \lesssim 0.36$ for $S=\frac{7}{2}$,
$0.275 \lesssim J_2 \lesssim 0.33$ for $S=\frac{9}{2}$, and
$0.265 \lesssim J_2 \lesssim 0.31$ for $S=\frac{11}{2}$.
For $S=\frac{5}{2}$, this is consistent with the range $0.3 \lesssim J_2 \lesssim 0.4$ estimated in Ref.~\onlinecite{Frederic_Mila_Spin5_2_PhysRevB.105.174402}. Although not shown here, it is evident that the edge state exists since the spin gap vanishes for all other possible sets of edge spins. Thus, for spin-$S$ cases, in these regions, the system forms a $(\frac{2S-1}{2}, \frac{2S+1}{2})$-type VBS PD state for $S_{\rm edge}=\frac{2S-1}{2}$ and $(\frac{2S+1}{2}, \frac{2S-1}{2})$-type one for $S_{\rm edge}=\frac{2S+1}{2}$. Similar to the $S=\frac{3}{2}$ case, the presence or absence of oscillations in the size dependence of the physical quantities allows us to distinguish between the PDC and PDIC phases. More details are given in the Supplemental Material. Furthermore, since the spin gap remains finite at the upper bound of the PDIC phase, it suggests that the region beyond the PDIC phase connects to a floating VBS phase with a finite spin gap. For these high $S$ systems, determining whether the FD phase emerges at even larger $J_2$ beyond the floating VBS phase is extremely challenging, and we have not yet been able to confirm it.

\begin{figure}[t]
\centering
\includegraphics[width=0.7\linewidth]{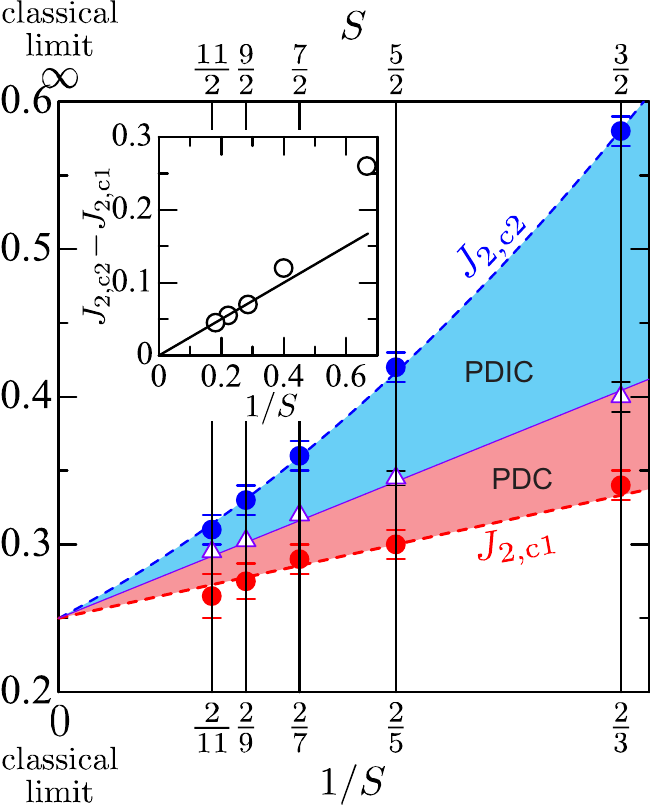}
\caption{
		Lower and upper bounds of the PD phase ($J_{2,{\rm c1}}$ and $J_{2,{\rm c2}}$, respectively) as a function of $1/S$. The boundary between PDC and PDIC phases is also plotted. Inset: The width of PD region, $J_{2,{\rm c2}}-J_{2,{\rm c1}}$, as a function of $1/S$, where the solid line represents $J_{2,{\rm c2}}-J_{2,{\rm c1}}=1/(4S)$.
}
\label{fig:PD}
\end{figure}

{\it Approaching the classical limit. ---}
Based on the results obtained, we can speculate on how the PD phase evolves as $S$ increases and approaches the classical limit $S \to \infty$. In Fig.~\ref{fig:PD}, we plot the lower and upper bounds of the PD phase, $J_{2,{\rm c1}}$ and $J_{2,{\rm c2}}$ respectively, as well as the boundary between the PDC and PDIC phases as functions of $1/S$. It is observed that both $J_{2,{\rm c1}}$ and $J_{2,{\rm c2}}$ approach $J_2 = \frac{1}{4}$ in the classical limit. Consequently, the boundary between the PDC and PDIC phases converges to $J_2=\frac{1}{4}$. In the classical model, $J_2 = \frac{1}{4}$ represents the boundary between the commensurate phase with $Q = \pi$ and the incommensurate phase with $Q = \pm \cos^{-1}\left(\frac{1}{4J_2}\right)$. Therefore, these results appear to be quite reasonable. Additionally, as shown in the inset of Fig.~\ref{fig:PD}, the width of the PD phase region behaves as $J_{2,{\rm c1}} - J_{2,{\rm c2}} \propto \frac{1}{4S}$ in the large $S$ regime.

{\it Summary. ---}
We investigate the ground state properties of antiferromagnetic $J_1$-$J_2$ chains with half-integer spins ranging from $S=\frac{3}{2}$ to $S=\frac{11}{2}$ using the density-matrix renormalization group method. Our study focuses on calculating the dimer order parameter, spin gap, and string order parameter as a function of $\frac{J_2}{J_1}$ for those high-$S$ systems. We have confirmed that topological dimerized phase characterized by alternating $\frac{2S-1}{2}$ and $\frac{2S+1}{2}$ valence bonds appears in a finite $J_2$ range up to $S=\frac{11}{2}$. These topological shrinks inversely with $S$, converging to a single point at $\frac{J_2}{J_1}=\frac{1}{4}$ in the classical limit -- a critical threshold between commensurate and incommensurate orders. Additionally, we have extended the discussion of the Majumder-Ghosh state, previously noted only for $S=\frac{1}{2}$, showing its potential presence as a ground state in half-integer higher spin systems over a substantial range of $\frac{J_2}{J_1}$ values.

{\it Acknowledgements.---}
We thank Ulrike Nitzsche for technical support. This work is supported by
the SFB 1143 of the Deutsche Forschungsgemeinschaft, SRG/2020/001203 of SERB, India, and F.30-528/2020 (BSR) of UGC, India.

\bibliography{highSJ1J2}

\end{document}


\title{Supplementary Material for\\ ``Topological Phases in Half-Integer Higher Spin $J_1$-$J_2$ Heisenberg Chains''}

\author{Sahinur Reja$^1$ and Satoshi Nishimoto$^{2,3}$}

\address{
$^1$ Department of Physics, Jadavpur University, Kolkata 700032, West Bengal, India\\
$^2$ Department of Physics, Technical University Dresden, 01069 Dresden, Germany\\
$^3$ Institute for Theoretical Solid State Physics, IFW Dresden, 01069 Dresden, Germany
}

\date{\today}

\begin{abstract}
Boundary between partially-dimerized commensurate and incommensurate phases\\
Finite-size scaling analysis of the spin gap for the floating VBS and fully-dimerized phases
\end{abstract}

\maketitle

\section{Boundary between partially-dimerized commensurate and incommensurate phases}

As discussed in the main text, the (partially dimerized) PD phase is further classified into PDC (PD-commensurate) and PDIC (PD-incommensurate) types. In our analyses, the PDC phase is characterized by the absence of oscillations, whereas the PDIC phase exhibits oscillations in the size-dependence of physical quantities. The boundaries between the PDC and PDIC phases are estimated as $J_2 = 0.345 \pm 0.005$ for $S = 5/2$, $J_2 = 0.320 \pm 0.005$ for $S = 7/2$, $J_2 = 0.3025 \pm 0.0025$ for $S = 9/2$, and $J_2 = 0.295 \pm 0.0025$ for $S = 11/2$.

\begin{figure}[tbh]
	\centering
	\includegraphics[width=0.8\linewidth]{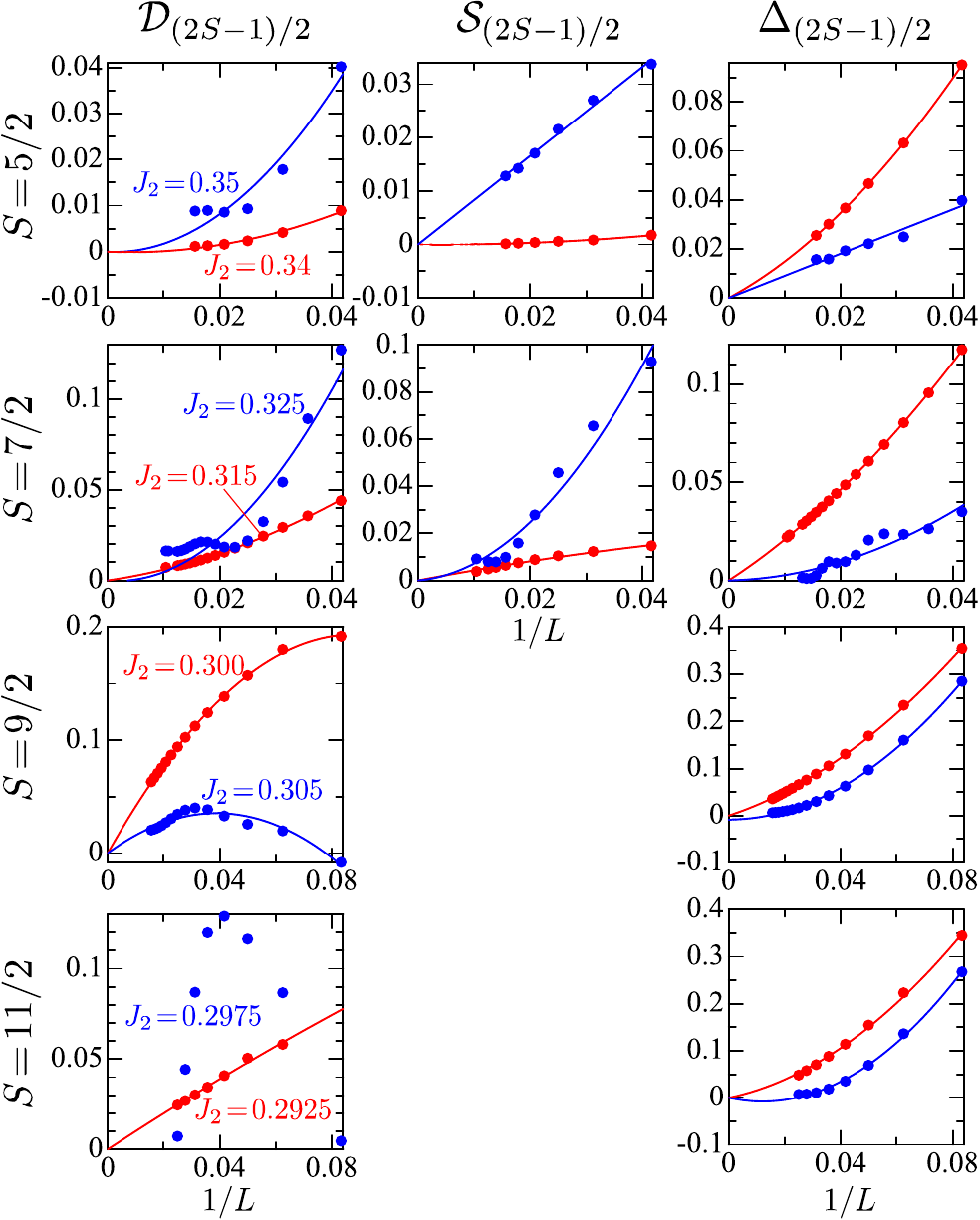}
	\caption{
		Size-dependence of the physical quantities: the dimer order parameter $\mathcal{D}{S{\rm edge}}$, the string order parameter $\mathcal{S}{S{\rm edge}}$, and the spin gap $\Delta_{S_{\rm edge}}$ around boundary between PDC and PDIC phases for $S=5/2$, $S=7/2$, $S=9/2$, and $S=11/2$. The thermodynamic limit value for each quantity is subtracted and the finite-size scaling is depicted with solid lines.
	}
	\label{fig:PDC_PDIC}
\end{figure}

To illustrate this, we present the size-dependence of the physical quantities: the dimer order parameter $\mathcal{D}{S{\rm edge}}$, the string order parameter $\mathcal{S}{S{\rm edge}}$, and the spin gap $\Delta_{S_{\rm edge}}$ for $S=5/2$, $S=7/2$, $S=9/2$, and $S=11/2$ in Fig.~\ref{fig:PDC_PDIC}. To clearly demonstrate the presence or absence of oscillations, we subtract the thermodynamic limit value for each quantity and depict the finite-size scaling with solid lines.

In all plots, when $J_2$ is less than the boundary value, the data points align smoothly with the scaling line. Conversely, when $J_2$ exceeds the boundary value, the data points oscillate around the scaling line. This clearly indicates that the boundary between the PDC and PDIC phases can be identified by the presence or absence of oscillations in the size scaling of physical quantities.

\section{Finite-size scaling analysis of the spin gap for the floating VBS and fully-dimerized phases}

As detailed in the main text, estimating spin gaps for the floating valence bond solid (VBS) phase ($0.58 \lesssim J_2 \lesssim 1.0$) and the fully-dimerized (FD) phase ($J_2 \gtrsim 1.0$) proves challenging. The first reason for this difficulty is the oscillatory behavior of the spin gap as a function of system size due to the incommensurate nature of the spin correlation functions. A second complicating factor is the increasing bandwidth of the magnon dispersion with larger $J_2$ values, which exacerbates the size-dependence of the spin gap. Conversely, the extrapolated spin gap size in the thermodynamic limit appears to decrease.

To address these challenges, we perform two types of finite-size-scaling analyses to analyze the results complementarily. Initially, we compute the spin gap, $\Delta_{S_{\rm edge}}$, for system sizes ranging from $L=12$ to $L=180$, in increments of four sites. In the first scaling analysis, we employ a polynomial fit of $\Delta_{S_{\rm edge}}$ against $1/L$, aiming to estimate the spin gap in the limit of $L \to \infty$. The characteristic oscillations and pronounced size-dependence are evident, particularly at larger values of $J_2$, as illustrated in Fig.~\ref{fig:fig:scalingS32}.

In a subsequent scaling analysis, we extrapolate the size by a two-step procedure: Initially plotting $\Delta_{S_{\rm edge}}$ from $L = L_1$ to $L = 180$ against $1/L$, and applying a linear fit to derive the value at $L \to \infty$, denoted as $\Delta_{1, S_{\rm edge}}(L_1)$. We then systematically vary $L_1$ from $L=12$ to $L=144$ and perform scaling for each $L_1$. This procedure yields a set of values for $\Delta_{1, S_{\rm edge}}(L_1)$, which we subsequently plot and fit linearly against $1/L_1$. As shown in Fig.~\ref{fig:fig:scalingS32}, while oscillations persist, the size dependence of $\Delta_{1, S_{\rm edge}}(L_1)$ is significantly reduced compared to the original measurements. Nevertheless, achieving reliable size scaling remains challenging at $J_2 = 1.20$ and $1.50$ when setting $S_{\rm edge} = 1/2$ or $1$, suggesting that edge states, possibly indicative of symmetry breaking localized at the system's edges, might emerge in the FD phase. However, the size scaling stabilizes most effectively in the FD phase when setting $S_{\rm edge} = 3/2$.

\begin{figure}[tbh]
	\centering
	\includegraphics[width=0.8\linewidth]{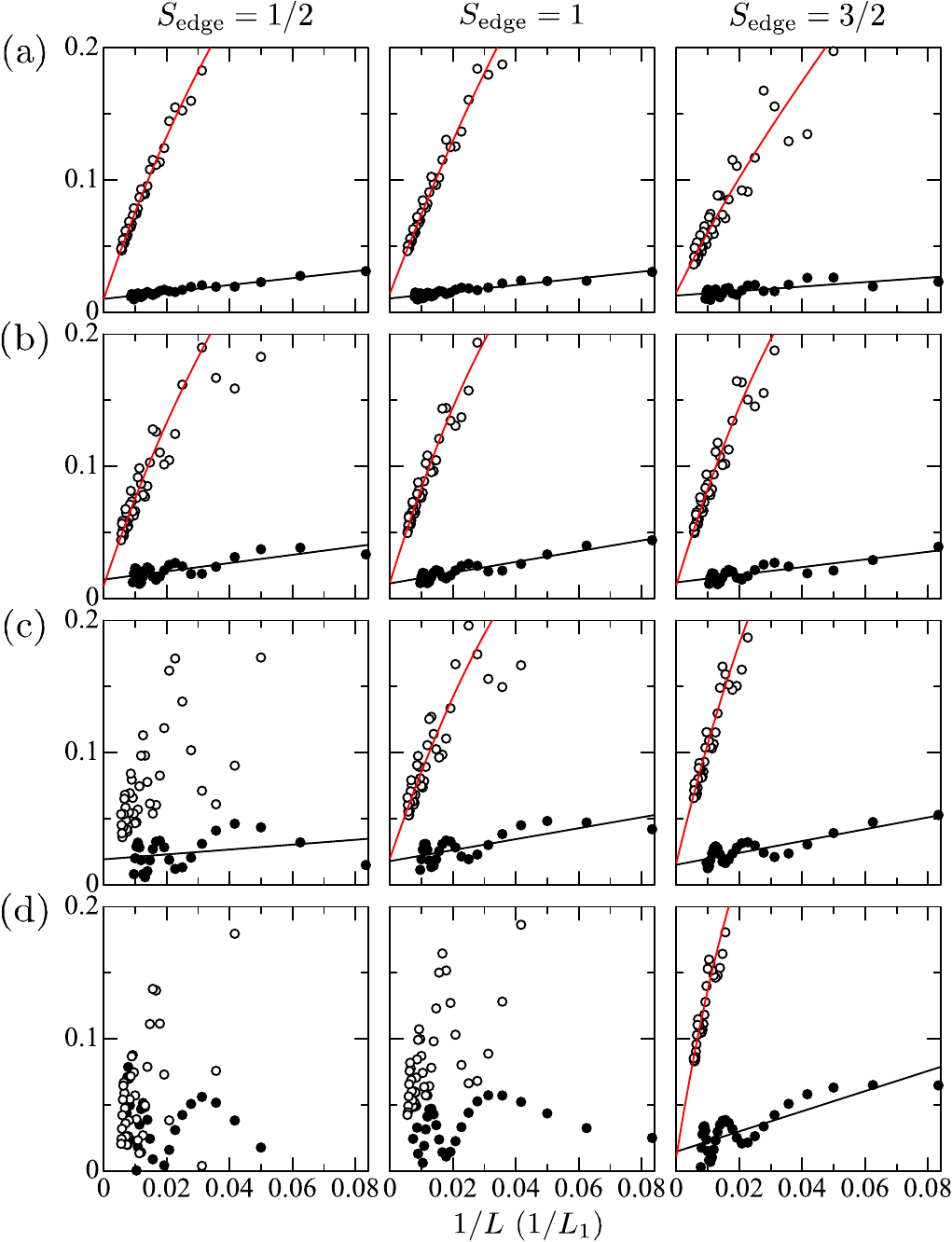}
	\caption{
		Finite-size-scaling analysis of the spin gap for $S=3/2$ at (a) $J_2=0.80$, (b) $1.00$, (c) $1.20$, and (d) $1.50$, considering edge spins $S=1/2$, $1$, and $3/2$. The open circles denote the bare data $\Delta_{S_{\rm edge}}$ and filled circles denote the refined data $\Delta_{1,S_{\rm edge}}(L_1)$ (see text).
	}
	\label{fig:scalingS32}
\end{figure}

\bibliography{highSJ1J2.bib}